\documentclass[10pt, conference]{IEEEtran}
\usepackage{geometry}
\usepackage[T1]{fontenc}
\pdfoutput=1
\IEEEoverridecommandlockouts
\makeatletter
\def\endthebibliography{%
	\def\@noitemerr{\@latex@warning{Empty `thebibliography' environment}}%
	\endlist
}
\makeatother

\IEEEoverridecommandlockouts

\usepackage{booktabs} 
\usepackage{cite}

\usepackage{setspace,multirow}
\usepackage{algpseudocode}
\usepackage{mathrsfs}
\usepackage{amsmath,amsfonts,amssymb,amsthm}
\usepackage{algorithm}
\usepackage{graphicx}
\usepackage{array}
\usepackage{caption}
\usepackage{makecell}

\usepackage[caption=false,font=footnotesize]{subfig}

\usepackage{epstopdf}
\usepackage[hyphens]{url}
\usepackage[dvipsnames]{xcolor}
\usepackage[normalem]{ulem}

\newtheorem{proposition}{Proposition}

\newtheorem{definition}{Definition}

\algrenewcommand\algorithmicrequire{\textbf{Input:}}
\algrenewcommand\algorithmicensure{\textbf{Output:}}

\newcommand{\bX}{\mathbf{X}}

\newcommand{\Xcal}{\mathcal{X}}

\newcommand{\Scal}{\mathcal{S}}
\newcommand{\Pcal}{\mathcal{P}}

\newcommand{\bx}{\mathbf{x}}

\newcommand{\balpha}{\boldsymbol{\alpha}}

\renewcommand{\IEEEQED}{\IEEEQEDopen}

\renewcommand{\IEEEQED}{\IEEEQEDopen}

\IEEEoverridecommandlockouts\IEEEpubid{\makebox[\columnwidth]{ 979-8-3503-1090-0/23/\$31.00 ~\copyright~2023 IEEE \hfill} \hspace{\columnsep}\makebox[\columnwidth]{ }}

\begin{document}
	
	\title{{On-board Change Detection for Resource-efficient Earth Observation with LEO Satellites}} 


\author{Van-Phuc Bui\IEEEauthorrefmark{1}, Thinh Q. Dinh\IEEEauthorrefmark{2}\IEEEauthorrefmark{4}, Israel Leyva-Mayorga\IEEEauthorrefmark{1}, Shashi Raj Pandey\IEEEauthorrefmark{1}, Eva Lagunas\IEEEauthorrefmark{3}, Petar Popovski\IEEEauthorrefmark{1}, \\ \IEEEauthorblockA{\IEEEauthorrefmark{1}Department of Electronic Systems, Aalborg University, Denmark (\{vpb, ilm, srp, petarp\}@es.aau.dk)\\
 \IEEEauthorrefmark{2}University of Information Technology, Ho Chi Minh City, Vietnam (thinhdq@uit.edu.vn)\\ \IEEEauthorrefmark{4}Vietnam National University, Ho Chi Minh City, Vietnam\\
 \IEEEauthorrefmark{3}University of Luxembourg, Luxembourg (eva.lagunas@uni.lu)
 }\thanks{This work was supported by the Villum Investigator Grant ``WATER'' from the Velux Foundation, Denmark. The work of Eva Lagunas has received funding from the Luxembourg National Research Fund (FNR) under the project SmartSpace (C21/IS/16193290).}}

	\maketitle

\begin{abstract}
The amount of data generated by Earth observation satellites can be enormous, which poses a great challenge to the satellite-to-ground connections with limited rate. This paper considers problem of efficient downlink communication of multi-spectral satellite images for Earth observation using change detection. The proposed method for image processing consists of the joint design of cloud removal and change encoding, which can be seen as an instance of semantic communication, as it encodes important information, such as changed multi-spectral pixels (MPs), while aiming to minimize energy consumption.  It comprises a three-stage end-to-end scoring mechanism that determines the importance of each MP before deciding its transmission. Specifically, the sensing image is (1) standardized and passed through a high-performance cloud filtering via the Cloud-Net model, (2) passed to the proposed scoring algorithm that uses Change-Net to identify MPs that have a high likelihood of being changed, compress them and forward the result to the ground station, and (3) reconstructed at ground gateway based on reference image and received data. The experimental results indicate that the proposed framework is effective in optimizing energy usage while preserving high-quality data transmission in satellite-based Earth observation applications.
\end{abstract}


\section{Introduction} \label{sec:Intro}
Remote sensing satellites are vital for environmental monitoring as they swiftly provide comprehensive coverage  of targeted regions, allowing for land use surveys, urban studies, and hazard management through image acquisition \cite{poursanidis2017remote}. 
The application of high-resolution sensors results in large volumes of sensor data, necessitating significant communication resources and on-board data storage capacity for transmitting data to ground-based end-users. However, the conventional approach of transmitting captured images to the ground for analysis and distribution is no longer practical due to the vast amounts of data generated by Earth observation satellites, prompting the need for novel processing techniques. 
For instance, the Sentinel-2 system acquires an extensive amount of data (2.4 Terabits per day) for transmitting to the terrestrial gateway, with each surface location being captured at periodic intervals of five days \cite{sentinel2}. While forwarding the complete dataset can be advantageous for the accurate detection of any alterations or anomalies, it is not regarded as efficient in terms of data storage and transmission capabilities.  Consequently, numerous forthcoming satellite missions aim to address memory and bandwidth limitations by implementing on-board processing operations that allow for data processing transfer from the ground segment to the space segment. The adoption of the new processing workflow is projected to significantly improve the efficiency of downlink data transmission leading to a decrease in the required transmission resources. Furthermore, by selecting solely the pertinent information to transmit, both the memory and communication costs of the on-board platform can be reduced \cite{trautner2010ongoing}.
There has been substantial interest in managing data in small Low Earth orbit (LEO) satellites due to their diminutive size, low power consumption, brief development cycle, and economic nature. It is important to note, however, that the communications capabilities of these  LEO satellites are limited, in contrast to their large captured data on a daily basis.
\begin{figure}[t]
	\centering
	\includegraphics[width=0.45\textwidth]{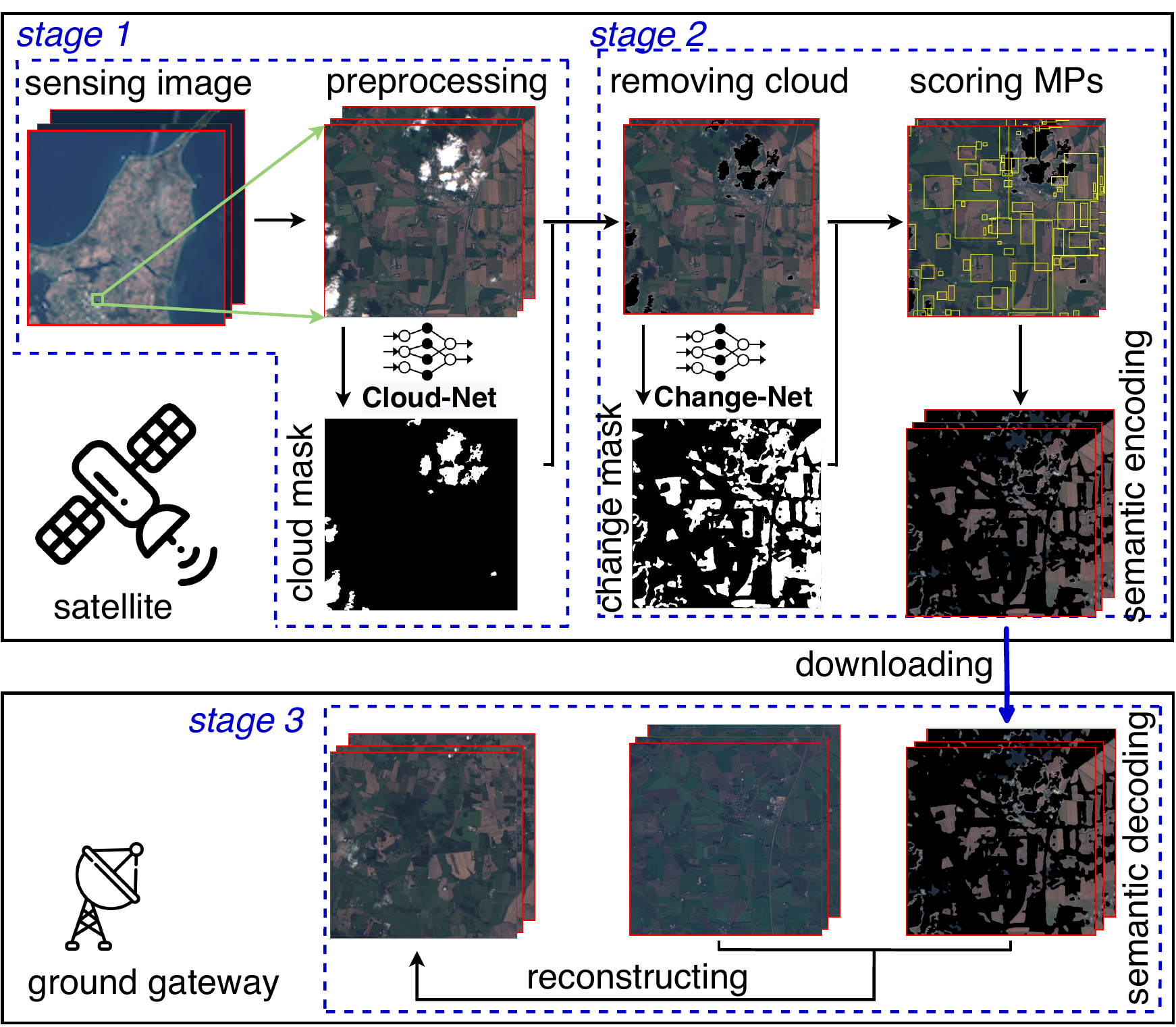}
	\vspace*{-0.1cm}
	\caption{Proposed end-to-end scoring architecture.}
	\label{fig_system}
	\vspace*{-0.8cm}
\end{figure}

With the latest developments in remote sensing applications based on artificial intelligence, the attention for machine learning grows steadily also for space information networks. 
Deep learning-based approaches have exhibited remarkable efficacy across numerous image-processing applications, and scientists working  in remote sensing have investigated the potential utility of these techniques in the context of Earth observation. Numerous Fully Convolutional Neural Networks (FCNNs) have been implemented for the purpose of semantic image segmentation, with many of them utilizing an encoder-decoder structure akin to that of the U-Net architecture \cite{ronneberger2015u}. The effectiveness of U-Net in delineating clouds in Landsat 8 imagery, aided by automatically generated Ground Truths (GTs), has been firmly established \cite{mohajerani2018cloud}. The domain of change detection has its origins in the early stages of aerial imaging \cite{hussain2013change}. Convolutional Neural Networks (CNNs), a category of machine learning algorithms specialized in image analysis, have been employed for image comparison in diverse settings \cite{zagoruyko2015learning}. 
It is immediate to see that such predictor cannot achieve perfect accuracy and, in fact, 
 it has been observed that change detection accuracy is in the range of $20-92\%$ even using all spectral bands for computing \cite{9740122}.
In addition, resource allocation in Space Information System is studied in  \cite{WangShengZhuangJSAC2018,WangShengYeArvix2019}. Scheduling problems that maximized the sum priorities of the successfully scheduled tasks  in \cite{WangShengZhuangJSAC2018} or the sum network utility \cite{WangShengYeArvix2019} have been considered. In those works, it is assumed that all satellites' collected information needs to be transmitted. 

The aforementioned works have primarily either focused on improving the performance of noise removal, such as clouds, or enhancing the accuracy of change detection methods (at the ground after receiving whole satellite's data), without fully considering the importance of transmitting critical information contained within noisy sensing images to the gateway.  Multi-spectral images are particularly susceptible to noise, such as cloud cover, which can result in the loss of useful information, including changing MPs. Therefore, the selection of pertinent MPs for downlink is crucial in conserving transmission resources at the satellite, as well as processing and storage space at both gateway and satellites. Consequently, this paper presents an end-to-end system that utilizes deep learning-based architecture to process and transmit valuable information from noisy sensing images captured by satellites. Our contributions are listed as follows:
\begin{itemize}
    \item To the best of our knowledge, this is the first study to examine an end-to-end change extraction system for satellite-based Earth observation. This system can help reduce energy required for processing and transmitting encoded information, thereby enabling more efficient information storage.
    \item We propose an architecture designed to remove cloud and assess the significant role of each MP in sensing images through scoring algorithm, enabling the selection of only essential MPs for encapsulation and transmission. 
    \item The proposed algorithm is evaluated through numerical simulations that utilize standard datasets and real-world data obtained from Sentinel-2 about North Jutland Region, Denmark. The results demonstrate that our system can effectively reduce the amount of encoding and transmission data required while preserving critical information.
\end{itemize}


\section{System Model and Problem Formulation} 

\label{sec:model_formulation}
\subsection{System Model}

We consider a satellite system as illustrated in Fig.~\ref{fig_system}, where there is a LEO satellite communicating with a gateway deployed at ground level, which is directly connected to a server. The satellite is in charge of capturing multi-spectral images and is equipped with an AI-module intended for image feature extraction. These features, such as potential pixel changes and their associated coordinates, must be transmitted within the limited time period when the satellite and gateway are connected. Consider two coregistered multi-spectral satellite images taken at different time instances $\bX^{t_0} =  \{x^{t_0}(i,j,k)| 1\leq i\leq H, 1\leq j \leq W, 1\leq k\leq D\}$ and $\bX^{t_1} =  \{x^{t_1}(i,j,k)| 1\leq i\leq H, 1\leq j \leq W, 1\leq k\leq D\}$ of size $H\times W \times D$, where $H$, $W$, and $D$ are the size of height, width, and spectral bands, respectively.  $\bX^{t_0}$ is the reference image taken at the reference time $t_0$ and $\bX^{t_1}$ is the newly observed image taken at time $t_1$. 

Our objective is to communicate the changed multi-spectral pixels (MPs) from the sensing image $\bX^{t_1}$ captured by the satellite to Earth while minimizing the data transmitted to ground and the energy consumption of the entire process. 
Specifically, given $\Xcal^{t_1}=\{\bx_{ij}^{t_1}\}$ as a set of available MPs in image $\bX^{t_1}$, e.g., $\bx_{ij}^{t_1}$ could be represented by the MP $(i,j)$  including all spectral bands, we need to select a subset $\Pcal$ MPs among $\Xcal^{t_1}$ $(|\Pcal|\leq |\Xcal^{t_1}|)$. The reference pixel set $\Xcal^{t_0} = \{\bx_{ij}^{t_0}\}$ of the reference image $\bX^{t_0}$ is processed as analogous. We define $\mathcal{S} = \{s_{ij} \} $ as the accurate change map where $s_{ij} \in \{0,1\} $ represents the changed/unchanged MP. In particular, $s_{ij} = 1$ if there is a pixel change at location $(i,j)$, and vice versa. 
\begin{definition}
Pixel change is defined as the temporal variation occurring of coregistered images at a specific location, including changes in land use, urban coverage, deforestation, and other similar types of deviation~\cite{8418840}.
\end{definition}

\subsection{Energy consumption}
In our analysis, we evaluate the energy utilization of the satellite by calculating the combined energy consumption associated with image compression and communication. These factors are greatly influenced by the data capacity transmitted between the satellite and the gateway.

We denote $\alpha_{ij} \in \{0,1\}$ is a binary decision variable with $\balpha \triangleq \{\alpha_{ij}\}$, where $\alpha_{ij} =1$ if we select the MP $\bx_{ij}^{t_1}$ as a candidate of $\Pcal$. As a consequence, the transmission set $\Pcal$ could be construct as 
\begin{IEEEeqnarray}{lll}
    \Pcal \triangleq \{\alpha_{ij}\bx_{ij}^{t_1}\}, \quad \forall \alpha_{ij}, \bx_{ij}^{t_1}.
\end{IEEEeqnarray}
Then, the total amount of data $C(\balpha)$ to transmit the set $\Pcal$ is 
\begin{IEEEeqnarray}{lll}
    C(\balpha) = \sum_{\alpha_{ij}\in\balpha} D\alpha_{ij}.
\end{IEEEeqnarray}
The model to calculate the satellite's energy spending for processing tasks captures the most relevant CPU parameters \cite{mao2017survey}.  The model establishes a direct relationship between the energy consumed per clock cycle and the square of the CPU clock frequency $f_\text{CPU}$  multiplied by the effective capacitance coefficient, which is specific to the processor under consideration. Thereby, we have $E_\text{cycle}^\text{proc}= {P^\text{proc}(f_\text{CPU})}/{f_\text{CPU}},$ 
where $P^\text{proc}(f_\text{CPU})$ is the power consumption during processing at the maximum CPU frequency. Noting that the supplied power is linear with the number of processor cores $N_\text{CPU}$, the energy consumption to process data $E^\text{proc}(\balpha)$ is modeled as 
\begin{IEEEeqnarray}{lll}
    E^\text{proc}(\balpha) = C(\balpha) \hat{R}^\text{comp} E_\text{cycle}^\text{proc},
\end{IEEEeqnarray}
where $\hat{R}^\text{comp}= e^{\kappa\rho} - e^{\kappa}$ is the compressing complexity, which is defined as the number of CPU cycles to compress one bit of data by a compression ratio $\rho$ and the positive constant $\kappa$ depending on used compression algorithm \cite{leyva2022satellite}.

For communication energy, we consider an interference-free communication channel  affected by additive-white Gaussian noise (AWGN) with free space path loss, and the noise power is denoted by $\varpi^2$. Consequently, the signal-to-noise ratio (SNR) for the downlink satellite-to-ground link at a particular time $t$, represented by $\gamma_t$, is calculated as
\begin{IEEEeqnarray}{lll}
    \gamma_t = G_{\text{tx}}G_{\text{rx}}P_{\text{tx}}\left(\frac{c}{4\pi d_tf_c\varpi}\right)^2,
\end{IEEEeqnarray}
where $G_{\text{tx}}$ and $G_{\text{rx}}$ stand for the transmitter and receiver antenna gains. $P_{\text{tx}}$ is the transmission power at the satellite, $f_c$ is the carrier frequency, and $c=2.998\times 10^{8}$ is the speed of light. $d_t$ can be calculated due to satellite's predefined connection time to the gateway, orbit time, and satellite' altitude.
Once we know the SNR, a proper modulation and coding scheme is selected to achieve the reliable communication following the DVB-S2 system \cite{morello2006dvb}. We denote a given throughput $\bar{R}$ [bps] and $\gamma_\text{min}(\bar{R})$ as the minimum SNR to achieve a reliable rate $\bar{R}$, the set of ordered pairs is defined as $\mathcal{Q}_\text{DVB-S2} = \{(\bar{R}, \gamma_\text{min}(\bar{R})\}$. The selection of the transmission rate for selected set $\Pcal$ is determined by adapting the modulation and coding scheme in order to attain the maximum data rate for reliable communication. Let $\gamma_t$ be  minimum SNR experienced at the gateway at time $t$, when the transmission of the data is initiated. At this time instant, the rate is selected as
\begin{IEEEeqnarray}{lll}
    R_\Pcal = \max\big\{\bar{R} \in \mathcal{Q}_\text{DVB-S2}:&\ \gamma_t \geq  \gamma_{\min}(\bar{R})\big\}
\end{IEEEeqnarray}
The downlink energy consumption for download all data $\Pcal$ is therefore calculated as
\begin{IEEEeqnarray}{lll}
    E^\text{trans}(\balpha) = \frac{P_\text{tx}C(\balpha)}{R_\Pcal}.
\end{IEEEeqnarray}

Summing up, the total energy for processing and transmitting the selected data $\Pcal$ is derived as
\begin{IEEEeqnarray}{lll}\label{power_require}
   E^\text{tot}(\balpha) = E^\text{proc}(\balpha)  + E^\text{trans}(\balpha).
\end{IEEEeqnarray}
\subsection{Energy Efficient Data Downloading With {Change-Detection} Constraint}
In contrast to conventional approaches aimed at improving efficiency of the change detection algorithm that is applied to the sensed images at the satellite \cite{9740122}, our focus is centered on the reduction of the energy required to transmit all changed to the nearby gateway, which shifts the focus towards the communication aspects of the system. That is, all $x_{ij}$ that have change map $s_{ij}=1$ will be scheduled to the transmission set $\Pcal$. This is achieved by setting the values of the decision variable $\balpha$ as $\alpha_{ij}\geq s_{ij}, ~\forall \alpha_{ij}, s_{ij}$, which ensures that all the changed MPs are transmitted.
Accordingly, the optimization problem is to minimize the communication energy subject to change detection constraints, which is formulated as
\begin{IEEEeqnarray}{rlll} \label{glob_problem}
	\text{P1}: \min_{\balpha}& ~~E^\text{tot}(\balpha) \IEEEyessubnumber\label{glob_problem_a}\\
   \mathrm{s.t.}~~ & \alpha_{ij} \geq s_{ij} \IEEEyessubnumber\label{glob_problem_b}.
\end{IEEEeqnarray}

\begin{proposition}
The problem P1 achieves optimality when the constraints \eqref{glob_problem_b} satisfy $\alpha_{ij} = s_{ij}, \forall i = \{1,2,\dots,H \}, j= \{1,2,\dots, W\}$.  This is straightforward to prove due to the fact that the energy decreases monotonously as the number of transmitted pixels, identified by the number of non-zero entries in $\balpha$, decreases.
\end{proposition} 
Note that \eqref{glob_problem} can only be solved if the satellite knows the indexes of all the changed MPs beforehand but this is not the case in reality. Instead, it is necessary to implement an AI model at the satellite to estimate $s_{ij}$ and, hence, perform inference on the changed MPs, which in turn requires a reformulation of P1.

\subsection{{Approximate} Formulation of Data Downloading with Change Detectors}
Motivated by the above discussion, a practical solution for MP scheduling is developed in this paper where we would like to build a predictor which can recognize most of the change MPs. 
 We define $s_{ij}^p\in\{0,1\}$ as the predictor for $s_{ij}$ given by the selected change detection algorithm, which achieves an error level $\epsilon$.
 Then, P1 is reformulated as follows:
\begin{IEEEeqnarray}{rll}\label{approx_problem}
    \text{P2}: \min_{\{\alpha_{ij}\}}& \sum_{i=1}^H\sum_{j=1}^W \alpha_{ij} \IEEEyessubnumber \label{approx_problem_a}\\ 
   \mathrm{s.t.} \ &   \alpha_{ij} \geq s_{ij}^p, \IEEEyessubnumber\label{approx_problem_b}\\
   & \mathrm{Pr}(s_{ij}^p =0 | s_{ij} =1) \leq \epsilon,\IEEEyessubnumber\label{approx_problem_c}
\end{IEEEeqnarray}
where $\epsilon$ is the threshold that ensures a sufficient number of change MPs are selected. It is evident that when the value of epsilon is decreased, the criteria for identifying change MPs become more rigorous, necessitating the selection of additional pixels for transmission. 
As an illustration, when setting $\epsilon=0.05$, the problem \eqref{approx_problem} is devised to guarantee the selection and labeling of no less than $95\%$ of change MPs. Accordingly, we propose an algorithm to obtain an efficient suboptimal solution to the problem \eqref{approx_problem}.

\section{Preprocessing and Semantic Encoding}
In {the following, we describe the steps to detect change MPs in Earth observation applications, which include preprocessing to remove cloud cover and the use of a deep learning model to determine the likelihood that each specific pixel is changed from previous versions of the image}. {Algorithm~\ref{alg_global} summarizes these steps.} The aim is to ensure that the reconstructed images at the gateway are as precise as possible, thereby facilitating their accurate interpretation by human/domain experts in Earth Observation. 

\begin{algorithm}[t]
	\begin{algorithmic}[1] \footnotesize
		\protect\caption{Proposed algorithm for problem  \eqref{approx_problem}} 
		\label{alg_global}
		\global\long\def\algorithmicrequire{\textbf{Input:}}
		\Require Geometrical multi-spectral image pair $\bX^{t_0}$ and  $\bX^{t_1}$
		\global\long\def\algorithmicrequire{\textbf{Output:}}
		\Require The transmission set $\Pcal$, Reconstructed MP image $\hat{\bX}^{t_1}$, and Energy requirement for processing and transmission
        \Statex \textit{Preprocessing and Cloud Removing}
        \State Select process bands (R, G, B, Nir) for cloud removing and (R, G, B) for change detection
		\State Standardize $\bX^{t_0}$ and  $\bX^{t_1}$ 
        \State Cloud detection and removing using Cloud-Net
        \Statex \textit{Change Scoring and Semantic Encoding}
        \State Scoring MPs using Change-Net and update $\{s_{ij}^p\}$ as in \eqref{score_update}
        \State Perform threshold segmentation to get the transmission set $\Pcal = \{\bx^{t_1}_{ij}| s^p_{ij} = 1, i = 1,2,\dots,H, j=1,2,\dots, W\}$
        \State Update $\alpha_{ij} = s^p_{ij} \ \forall i, j$ and calculate required energy $E^\text{tot}(\balpha)$ as in \eqref{power_require}
        \end{algorithmic}
\end{algorithm}

\subsection{Preprocessing and Cloud Removing}
As part of the preprocessing stage, the following steps are undertaken: $(1)$ the selection of bands representative of the complete set of MP images; $(2)$ the adjustment of illumination levels to facilitate the detection (cloud and change) in the subsequent stage; and $(3)$ the elimination of cloud cover from the observed image. First, $N\ (N<D)$ useful bands are selected for processing  based on the principle that a group of appropriate bands (e.g., visible and near-infrared bands) could reduce the processing requirements for both training and execution phases. This is consistent with realizing cloud detection on the satellite with limited hardware and processing capabilities.
Before removing clouds, a radiometric correction procedure is implemented to rectify radiometric disparities among multi-spectral images stemming from diverse imaging conditions such as sun angle, light intensity, and atmospheric circumstances. The chosen approach is relative radiometric normalization, which is founded on the z-score method, which normalizes images to zero mean  and a unit standard deviation.
Given a MP pair, denoted by $\bx_{ij}^{t_0} = [x_1^{t_0}, x_2^{t_0},\dots,x_N^{t_0}]$ and  $\bx_{ij}^{t_1} = [x_1^{t_1}, x_2^{t_1},\dots,x_N^{t_1}]$. 
The radiometric relative normalization can be expressed as $  \hat{x}_n^{t_0} = ({x_n^{t_0}-\mu_{\bx_n^{t_0}}})/{\sigma_{\bx_n^{t_0}}^2};  \hat{x}_n^{t_1} = ({x_n^{t_1}-\mu_{\bx_n^{t_1}}})/{\sigma_{\bx_n^{t_1}}^2},$
where $\mu_{\bx_n^{t_k}}$ is the mean and $\sigma_{\bx_n^{t_k}}^2$ is the variance for band $n$ of image $\mathbf{X}^{t_k}, (k\in\{0,1\})$ . The radiometric correction would suppress the radiometric difference between multi-spectral images caused by different conditions.

\textit{Cloud removing:} In this work, the Cloud-Net model for identifying cloud regions in multi-spectral Sentinel-2 images based on U-Net using a combination of thresholding and deep learning techniques is deployed. 
This approach utilizes four spectral bands - Red, Green, Blue, and Near Infrared (RGBNir) - for both training and prediction, making it practical for use in satellite hardware for cloud detection. 
As each spectral band in Sentinel-2 encompasses a substantial number of pixels (approximately $10980 \times 10980$), it is necessary to segment them into smaller image patches using a cropping technique. This is achieved by dividing the selected spectral band image into $384 \times 384\times 4$ non-overlapping patches, with each of the four patches corresponding to the Red, Green, Blue, and Nir bands being combined to create a 4D input. To enhance the resilience of the methodology towards cloud patterns or similar patterns, the input patches underwent geometric transformations, such as horizontal flipping, rotation, and zooming.
{The output probability map is obtained via the use of a sigmoid activation function in the final convolution layer of the network. 
The cloud learning set consists of tuples $\mathcal{D}_\text{cloud}=\{(\mathbf{a}_{i}, z_{i}) \in (\mathcal{A}\times \mathcal{Z})|i = 1, \dots, M_\text{cloud})\}$ obtained
from an unknown joint distribution $P_{\mathcal{D}_\text{cloud}}$ over $\mathcal{A}\times \mathcal{Z}$, where $M_\text{cloud}$ is the total number of instances, and $z_{i}$ represents the corresponding label of the input $\mathbf{a}_{i}$. Our aim is to estimate a function $g(\mathbf{a}_{i}|\boldsymbol{\psi})$ that maps inputs $\mathbf{a}_{i}$ to outputs $z_i$, where $\boldsymbol{\psi}$ is a set of parameters that are optimized using the training set. To estimate $\boldsymbol{\psi}$, the Adam gradient descent method is utilized to implement the Cross Entropy loss function as 
\begin{IEEEeqnarray}{lll}
    \mathcal{L}(\hat{z}_{\boldsymbol{\psi},i},z_{i} ) = - \big(z_{i} \log(\hat{z}_{\boldsymbol{\psi},i}) 
    &+ (1-z_i)\log(1-\hat{z}_{\boldsymbol{\psi},i})\big),\IEEEeqnarraynumspace
\end{IEEEeqnarray}
with $\hat{z}_{\boldsymbol{\psi},i} = g(\mathbf{a}_{i}|\boldsymbol{\psi})$ is the output of the model.
The well-trained weights of the Cloud-Net model are then applied to the target image to generate a predicted cloud probability map. This map is binarized using a global threshold of $\gamma$ and multiplied with the image ${\bX}^{t_1}$ to produce the cloud-removed image $\Tilde{\bX}^{t_1}$ with $\Tilde{\Xcal}^{t_1} = \{\Tilde{\bx}^{t_1}_{ij}\}$ as a set of MPs in $\Tilde{\bX}^{t_1}$. }

\begin{figure*}[t]
\begin{minipage}{0.33\textwidth}
	\includegraphics[trim=0.cm 0.0cm 0.cm 0.1cm, clip=true, width=2.2in]{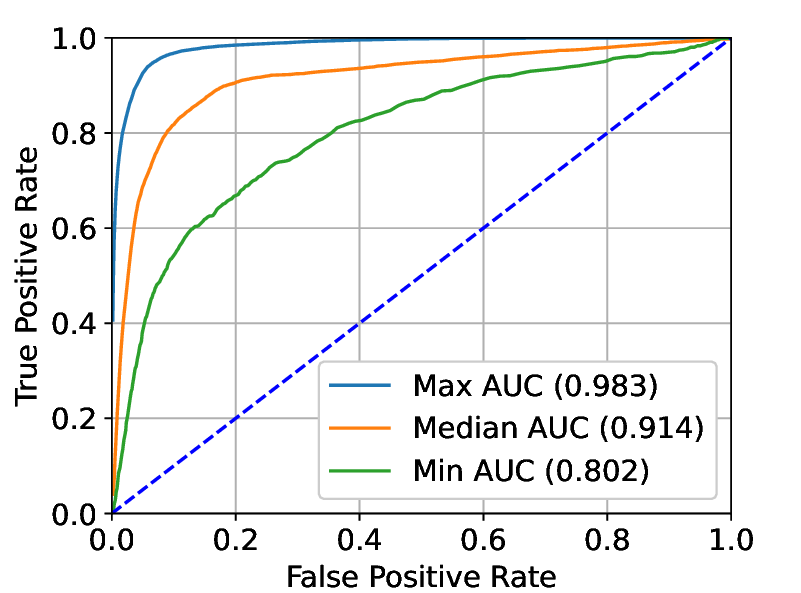} \\ 
	\vspace*{-0pt}
	\centering \fontsize{8}{8}{(a)}
	\vspace*{-5pt}
\end{minipage}
\begin{minipage}{0.33\textwidth}
	\includegraphics[trim=0.cm 0.0cm 0.cm 0.8cm, clip=true, width=2.2in]{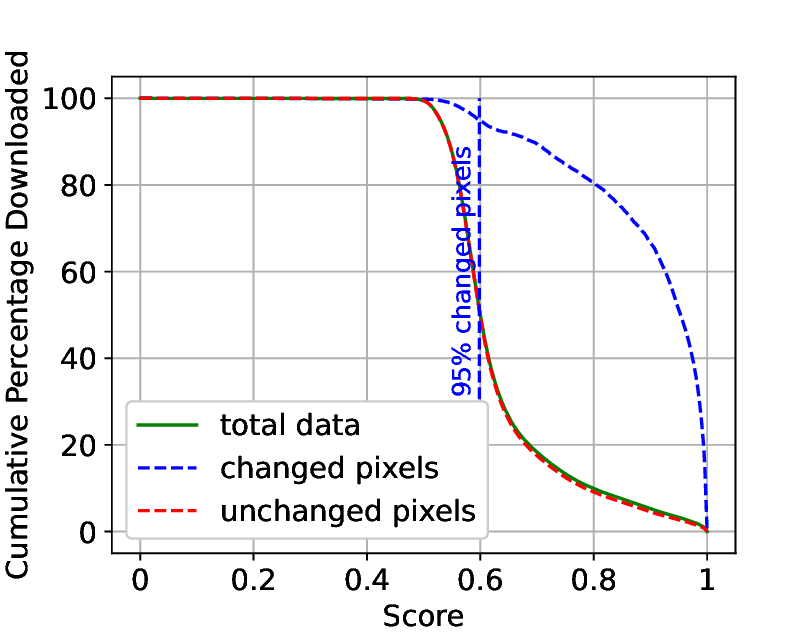} \\ 
	\vspace*{-0pt}
	\centering \fontsize{8}{8}{(b)}
	\vspace*{-5pt}
\end{minipage}
\begin{minipage}{0.33\textwidth}
	\includegraphics[trim=0.cm 0.0cm 0.cm 0.8cm, clip=true, width=2.2in]{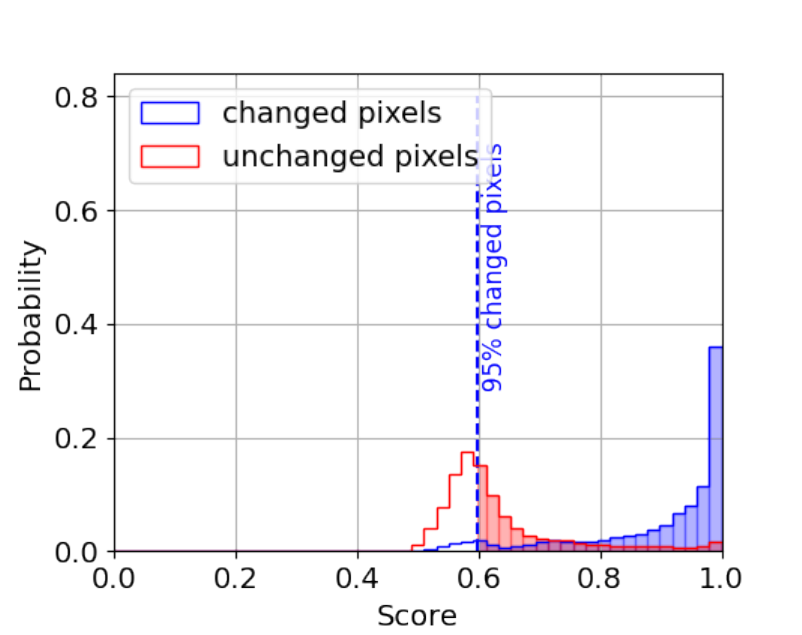} \\ 
	\vspace*{-0pt}
	\centering \fontsize{8}{8}{(c)}
	\vspace*{-5pt}
\end{minipage}
\caption{(a) ROC curve over change dataset (the AUC scores are shown in the legends). (b) Cumulative Percentage Downloaded vs score of multi-spectral image with median AUC. (c) Probability of changed and unchanged MPs vs score of multi-spectral image with median AUC.}
\label{fig_PDF}
\vspace*{-5pt}
\end{figure*}
\subsection{Change Scoring and Semantic Encoding}
In this stage, we have been provided with two multi-spectral images that have been geometrically aligned, have consistent lighting conditions, and have had cloud cover removed. Our goal is to propose a computationally efficient automatic change detection method for these two images that are practical for satellite applications. 
Motivated by \cite{ronneberger2015u}, the Change-Net based on U-Net model is developed to perform semantic segmentation of images.
We use a negative log-likelihood loss function to classify every MP in the observed images as either changed or not.  
{Specially, we are considering a change learning set consisting of tuples $\mathcal{D}_\text{change} = \{(\mathbf{b}_{i} , y_{i}) \in (\mathcal{B}\times \mathcal{Y})|i = 1, \dots, M_\text{change}\}$, with inputs $\mathbf{b}_{i}$, and labels $y_{i}$. With $\boldsymbol{\theta}$ as a set of learnable parameters to be optimized, the likelihood $p(\mathbf{b}_{i},\boldsymbol{\theta})$ is defined as the joint density of the observed data, which can be viewed as a function of $\boldsymbol{\theta}$ that maps any given input $\mathbf{b}_{i}$ to outputs $y_{i}$. By denoting $m$  as the mini-batch size, the parameter set $\boldsymbol{\theta}_\text{model}$ is then achieved through training process as
\begin{IEEEeqnarray}{lll}
   \boldsymbol{\theta}_\text{model} = \underset{\boldsymbol{\theta}}{\arg\max}\prod_{i=1 }^mp(y_i|\mathbf{b}_{i},\boldsymbol{\theta}).
\end{IEEEeqnarray}

In our case, the system output $\Bar{s}_{ij} \in [0,1]$ is defined as the change score of the corresponding cloud-removed MP $\Tilde{\bx}^{t_1}_{ij}$, which is the probability of $p(s_{ij}=1|\Tilde{\bx}^{t_1}_{ij},\boldsymbol{\theta}_\text{model}) = \xi(\boldsymbol{\theta}_\text{model}^T\Tilde{\bx}^{t_1}_{ij})$ with $\xi(\cdot)$ is the Log Softmax function. As $\Bar{s}_{ij}$ is close to $1$, it is likely that there is a change in location $(i,j)$. Thus, to make a good prediction, we would like to learn a scoring system such that
\begin{IEEEeqnarray}{rll}\label{score_update}
    s_{ij}^p = \mathbf{1}(\Bar{s}_{ij} \geq \tau),
\end{IEEEeqnarray}
where $\tau$ is the predefined threshold, derived experimentally to satisfy the constraint \eqref{approx_problem_c}.} Our proposed architecture is summarized in Algorithm 1.

\section{Numerical Results}
\begin{table}[!t]\vspace{-0.5cm}
\caption{Simulation Parameters}
\resizebox{8.9cm}{!} 
{\begin{tabular}{lr|lr}
		\hline
		Parameter & Value & Parameter & Value \\
		\hline
		Carrier frequency ($f_c$)          & 20 GHz      &  {Altitude} of satellites ($h$) \cite{sentinel2}          &   786 km    \\
            Processor frequency ($f_\text{CPU}$)          & 1.8 GHz      &   Compression factor $(\rho)$ &  5\\
            Power consumption          &      & Complexity of the image & \\
             for processing ($P^\text{proc}(f_\text{CPU})$)   & 10 W&  processing algorithm $(\kappa)$&  0.1\\
		Bandwidth   ($B$)       & 500 MHz      &  Noise power $(\varpi^2)$         & {-115 dB}     \\
            Satellite antenna gain $(G_\text{tx})$  & 32.13 dBi & Gateway antenna gain $(G_\text{rx})$ & 34.2 dBi\\
         Performance in change detection $(\epsilon)$ & 0.05   & Transmission power $(P_\text{tx})$ & 10 W      \\      
          Communications duration ( $T^{\mathrm{pass}}$ )  &   15 min   &  Orbital period &   100 min  \\   
		\hline\vspace{-0.5cm}
	\end{tabular}
}
\label{parameters}
\end{table}
\subsubsection{Implementation Details}
We evaluate our proposed algorithm in the Sentinel-2 system when the transmission process initiates upon the satellite's entry into the gateway's coverage area. This is a worst-case scenario for the energy consumption for communication. Other important parameters are included in Table \ref{parameters}.
We use the 38-Cloud dataset \cite{mohajerani2019cloud} for training and testing Cloud-Net, which covers 38 Landsat-8 scene images and their manually extracted pixel-level ground truths for cloud detection.
For Change-Net model, we conduct the Onera Satellite Change Detection dataset, publicly available on the IEEE-DataPort repository \cite{dataset_change_detection}, widely used in the literature to detect changes in multispectral images of Sentinel-2 satellite. 
The weights were assigned inversely proportional to the number of examples in each class in order to address class imbalances between the two categories (change versus no change). The features are preprocessed classically to facilitate training. The dataset was split into a shuffled and stratified training set and a test set, comprising $60\%$ and $40\%$ of the data, respectively.  Both Cloud-Net and Change-Net are trained on the ground and applied to satellites for further processing of sensing data with learning rate $10^{-3}$ and batchsize 10. Following training and evaluation on the dataset with annotated ground truth, we apply the trained algorithm to real-world data collected from a Sentinel-2 image to identify changes at the pixel level relative to reference images. These identified changed map $\Scal$ are subsequently labeled and encoded prior to being transmitted to the gateway.

\begin{figure*}[!thp]
\centering
\subfloat[Reference image]{\label{fig:1}{\includegraphics[width=0.23\textwidth]{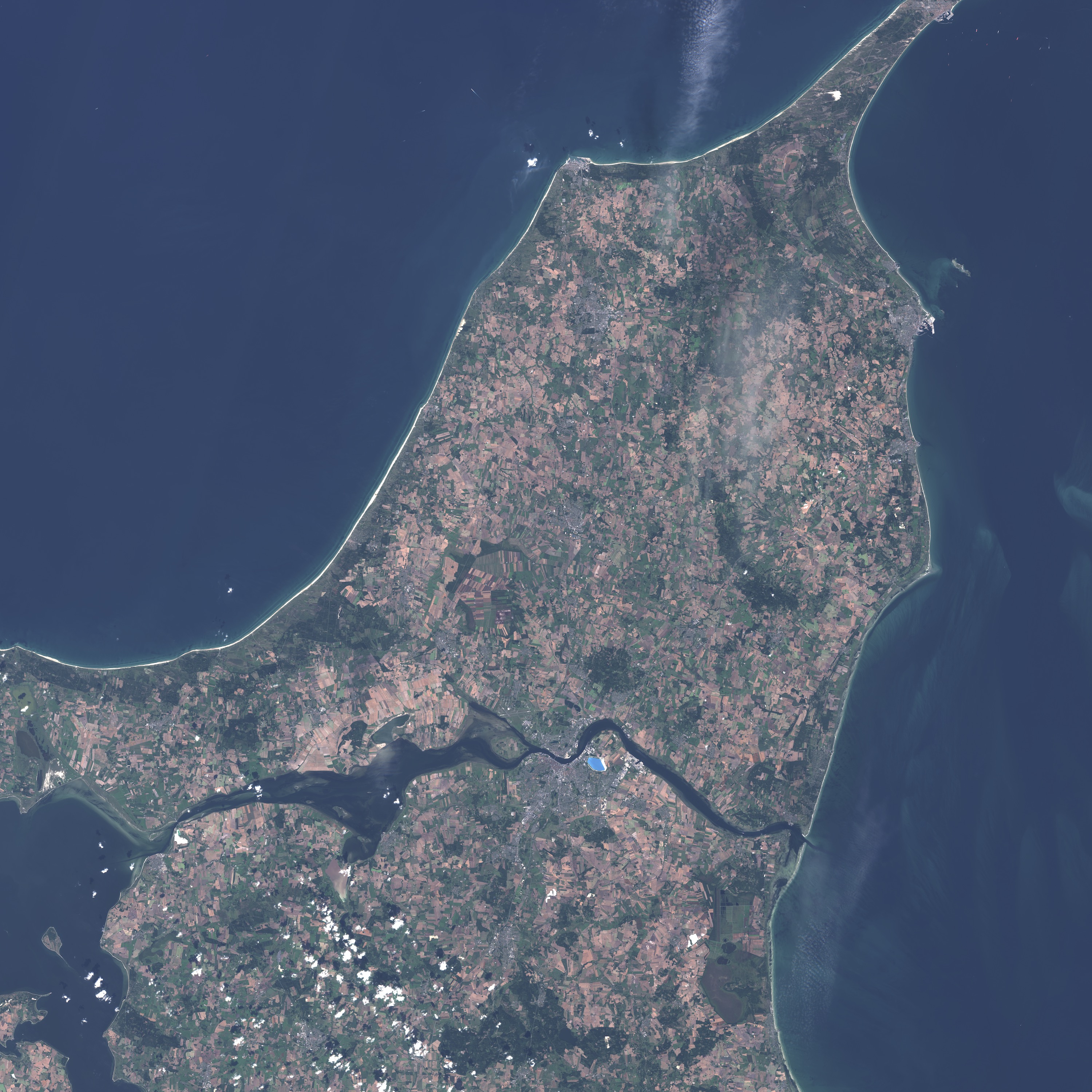}}}\hfill
\subfloat[Captured image (on 19/09/2022)]{\label{fig:2}{\includegraphics[width=0.23\textwidth]{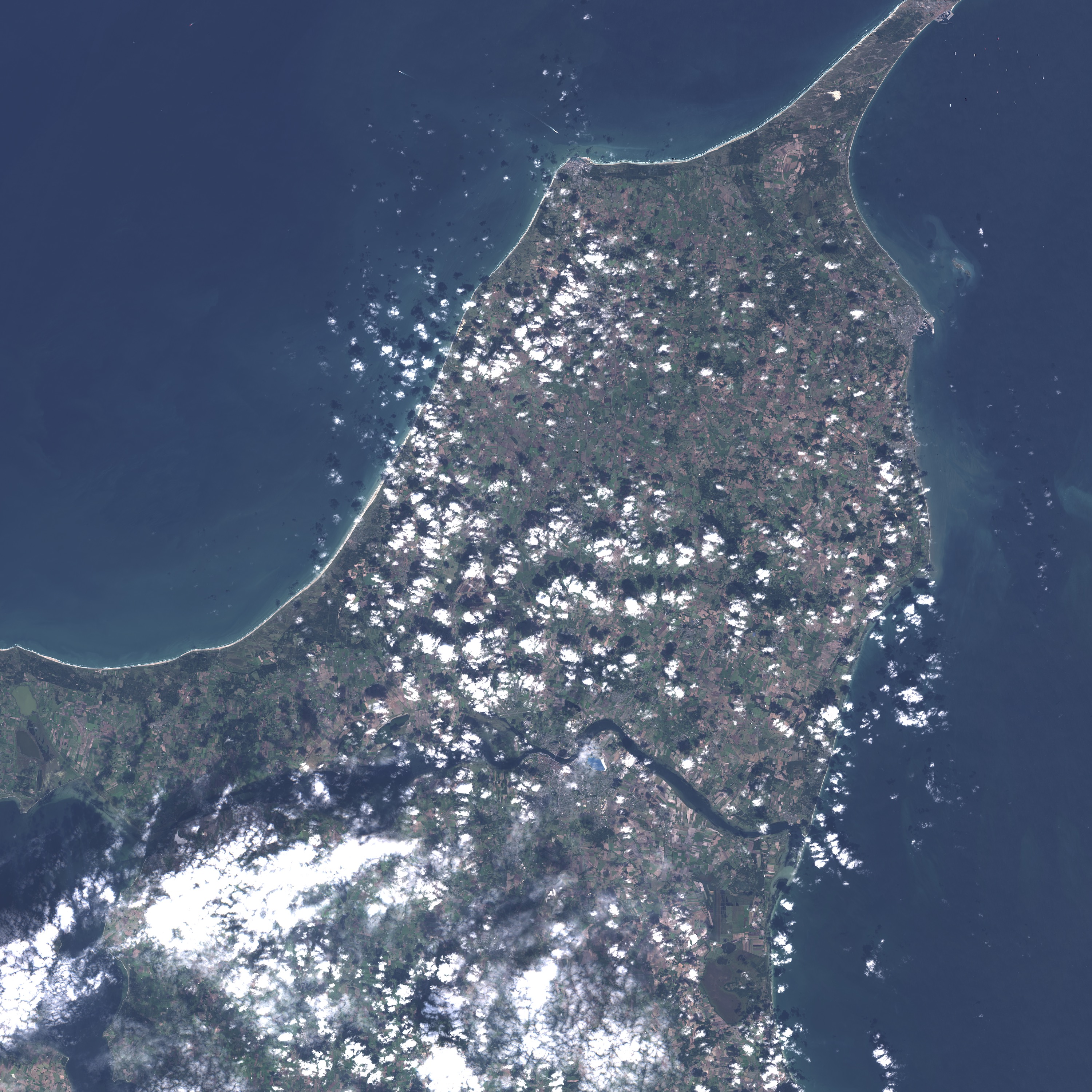}}}\hfill
\subfloat[Reconstructed image at Earth gateway (PSNR = 60.1 dB)]{\label{fig:3}{\includegraphics[width=0.23\textwidth]{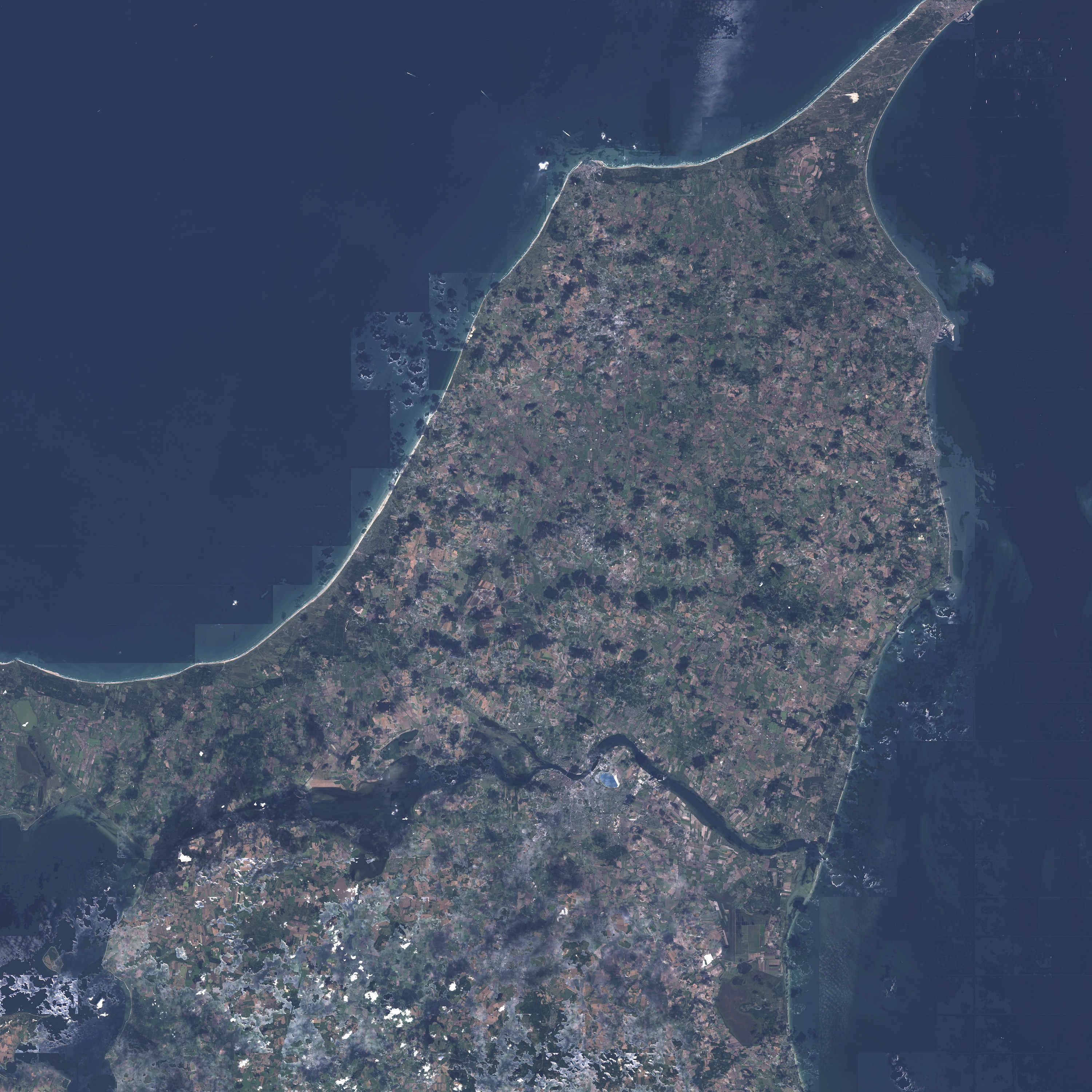}}}\hfill
\subfloat[Percentage of cloud and change prediction]{\label{fig:4}\hfill
{\includegraphics[width=0.4\textwidth]{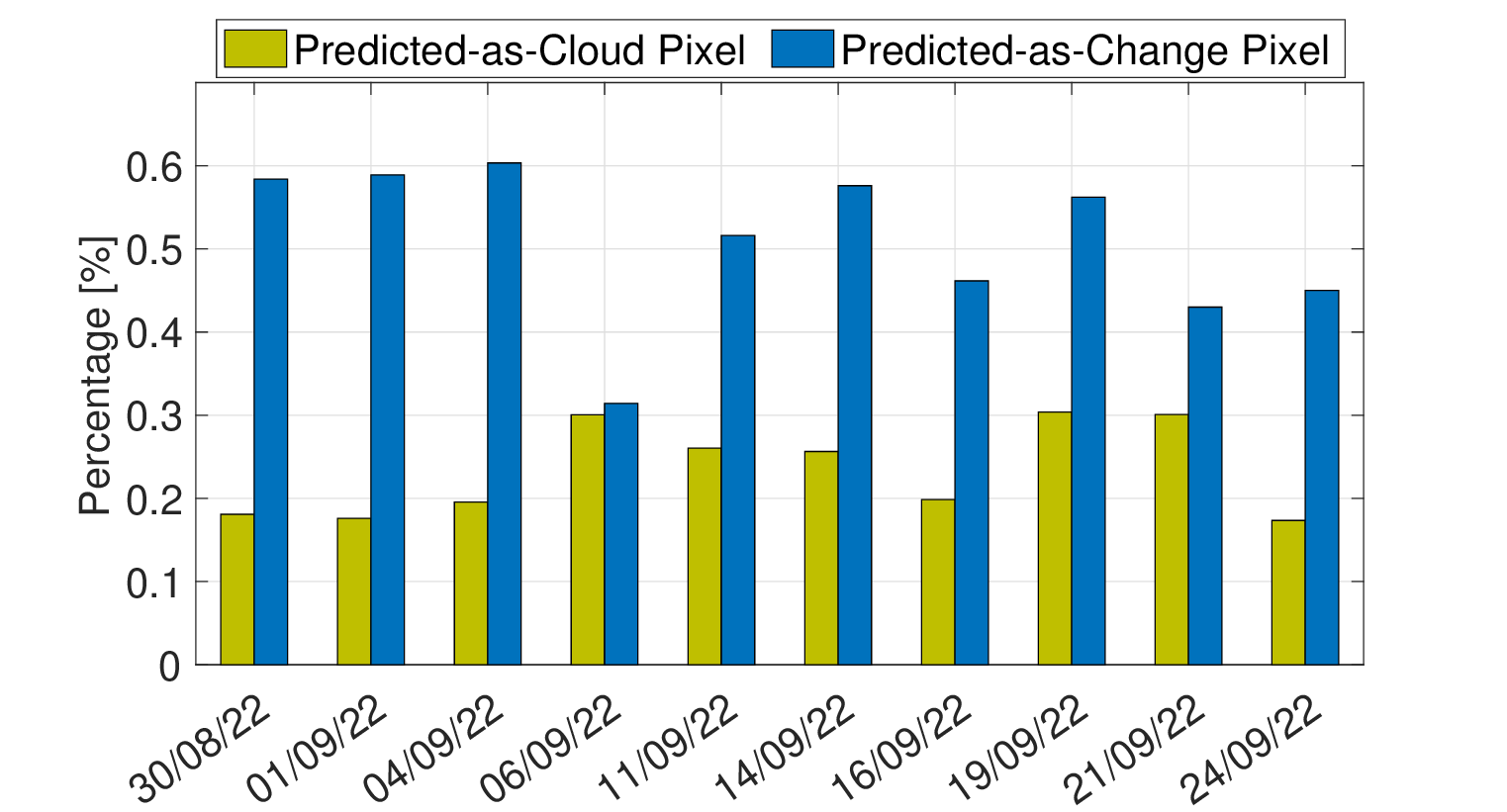}}}\hfill
\subfloat[Energy requirement for processing and transmission]{\label{fig:5}{\includegraphics[width=0.4\textwidth]{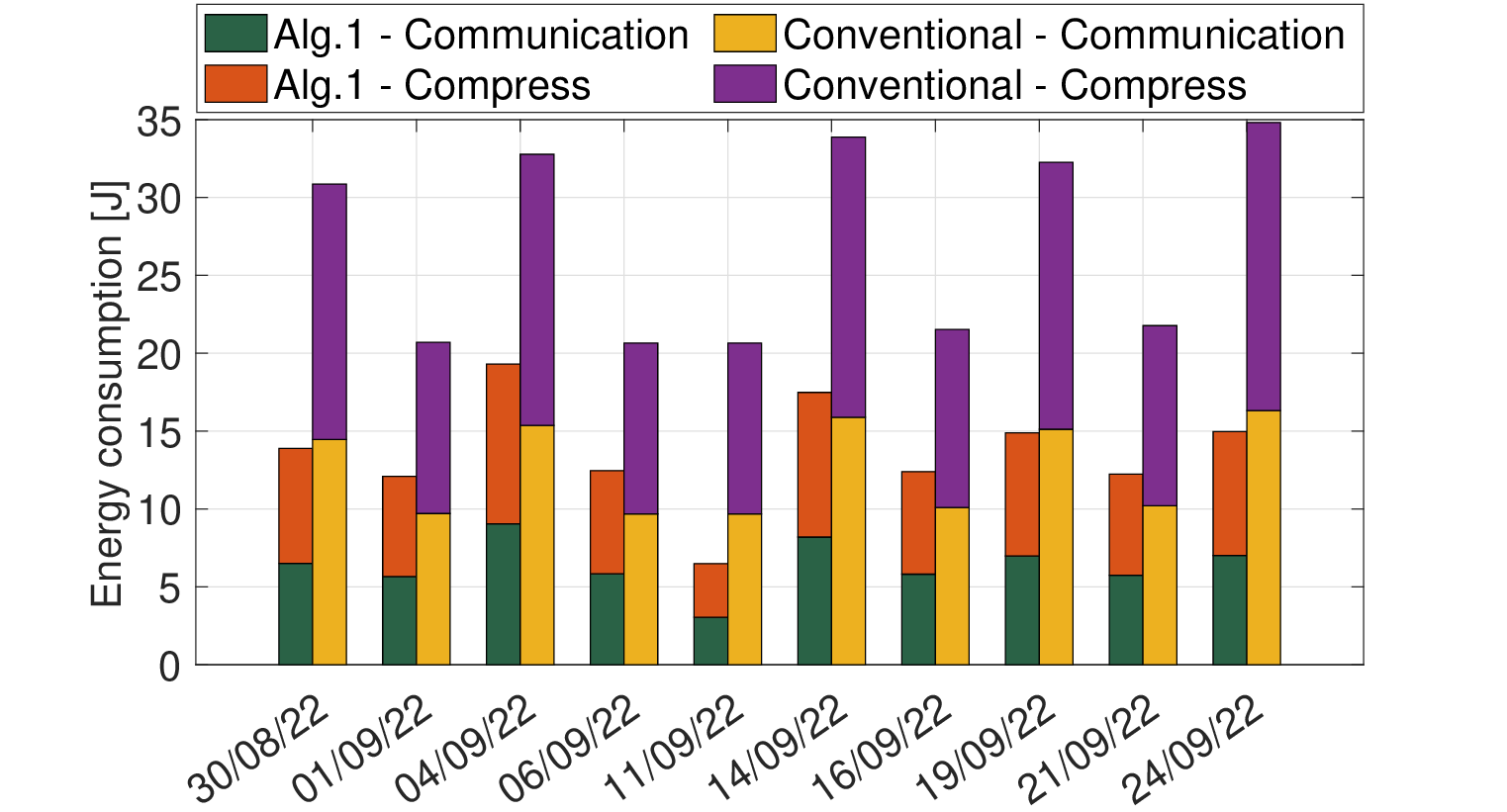}}}
\caption{The performance of proposed algorithm on the real Sentinel-2 data (the multi-spectral images were acquired in the North Jutland Region of Denmark during the period from 23/08/2022 to 24/09/2022, with the reference image captured on 23/08/2022 (figures (a), (b), and (c) are display using RGB channels).}
\label{fig_real_data}
\vspace{-0.5cm}
\end{figure*}

\subsubsection{Results}
We evaluate the  ROC of the change detection model with different thresholds. Herein, the True Positive Rate (TPR) and False Positive Rate (FPR) are respectively defined as    $ \text{TPR} ={\text{TP}}/({\text{TP}+\text{FN}}),$ and $\text{FPR} = {\text{FP}}/({\text{FP}+\text{TN}}).$
Based on ROC, the Area Under the ROC Curve (AUC) could be calculated, which measures how well predictions are ranked.

Fig.~\ref{fig_PDF}(a) plots the ROC of 3 representative images in the OSCD dataset with the highest, median, and lowest AUC levels, respectively. The efficacy of our trained Change-Net model can be readily observed, as it achieves a high level of performance with the maximum AUC of 0.98 and a minimum AUC of 0.80. Therefore, selecting the appropriate score threshold to guarantee the fulfillment of constraint (9) while simultaneously ensuring the selection of an adequate number of modified points can be considered a dependable approach. Fig.~\ref{fig_PDF}(b) and Fig.~\ref{fig_PDF}(c) depict the percentage of downloads accumulated and the probability of the image that has the median AUC score, respectively, while also presenting a threshold score of 0.6 to guarantee the selection of $95\%$ of modified MPs. Remarkably, MPs labeled 0 and 1 distribute overlap as in Fig.~\ref{fig_PDF}(c). Hence, to guarantee the selection of a sufficient number of altered MPs (at a rate of 95\%), we have labeled approximately 50\% of unchanged MPs with a value of 1, leading to the requirement of encoding and transmitting approximately 50\% of the entire sensing data. We also emphasize that our approach prioritizes the establishment of a threshold for labeling a specific portion of change MPs after training rather than evaluating the model's performance metrics over the test set.

To assess our architecture's efficacy, we incorporated real-world data from the Sentinel-2 system, taking into consideration both the quantity of reduced data and the computational required energy for processing and transmission. Remarkably, Fig.~\ref{fig_real_data}(a)-(c) illustrate a specific use case of images. The results indicate that Change-Net yields advantages  when the Peak signal-to-noise ratio (PSNR) between the reconstructed image (Fig.~\ref{fig_real_data}(c)) and the captured image (Fig.~\ref{fig_real_data}(b)) is 60.1 dB subsequent to cloud removal and without compression. In order to attain this outcome, Change-Net opted to select up to  60\% of MPs (as in  Fig.~\ref{fig_real_data}(d) on 04/09/2023) in order to ensure a sufficient selection of change MPs. Fig.~\ref{fig_real_data}(e) shows the required energy to compress and transmit data from the satellite. Compare to the conventional scheme (transmit the whole sensing image), our architecture saves at least  40\% in required energy, thereby directly saving both energy for data processing and transmission from satellites, as well as storage space and processing time at Earth gateways.

 \vspace*{-5pt}
\section{Conclusion}
This research examined the Earth observation scenario in order to enhance processing efficiency and minimize energy usage at edge devices during the transmission and storage of sensing data. To achieve this objective, we propose an effective deep learning-based scoring framework that automatically and encodes semantic MPs from multi-spectral sensing images. The numerical outcomes demonstrate that our end-to-end solution had superior data detection capability (95\%) and save at least 40\% energy consumption spent on data transmission.
\setstretch{0.85}
\vspace*{-5pt}
\bibliographystyle{IEEEtran}
\bibliography{Journal}
\vspace{-0.2cm}
\end{document}